\documentclass{openjournal}
\usepackage{natbib}
\usepackage{graphicx,amsmath,amssymb,amstext}
\usepackage{amsbsy,amsfonts,amsthm,color}
\usepackage[colorlinks,linkcolor=blue,citecolor=blue,urlcolor=blue ]{hyperref}
\usepackage[caption=false]{subfig}
\usepackage{float}
\begin{document}

\title{Gravitational Stability of Vortices in Bose-Einstein Condensate Dark Matter}
\author{Mark N Brook$^{1,2}$}
\author{Peter Coles$^*$$^{1,3}$}

\email{$^*$Peter.Coles@mu.ie}

\affiliation{$^{1}$ School of Physics \& Astronomy, University of Nottingham, Nottingham, NG7 2RD, UK.}
\affiliation{$^{2}$ The Institute of Cancer Research, 15 Cotswold Rd, Sutton SM2 5NG, UK.}
\affiliation {$^{3}$ Department of Theoretical Physics, Science Building, Maynooth University, Maynooth, Co. Kildare, Ireland.}


\begin{abstract}
We investigate a simple model for a galactic halo under the
assumption that it is dominated by a dark matter component in the
form of a Bose-Einstein condensate involving an ultra-light scalar
particle. In particular we discuss the possibility that if the dark
matter is in a superfluid state then a rotating galactic halo might
contain quantised vortices which would be low-energy analogues of
cosmic strings. Using known solutions for the density profiles of
such vortices we compute the self-gravitational interactions in such
halos and place bounds on the parameters describing such models,
such as the mass of the particles involved.
\end{abstract}


\section{Introduction}
\label{Introduction}

In the standard model of galaxy formation, the visible component of
a galaxy is supposed to be embedded in an invisible halo of
non-baryonic matter; see, for example, 
\cite{Jenkins:1997en} and \cite{Coles:2005yk}. This dark
component is further supposed to be {\em cold}, meaning that it is
usually assumed to consist of very heavy particles with very low
thermal velocities. However, it has been known for some time that
Cold Dark Matter (CDM) models have certain problems in reproducing
observable properties of galaxies, among them being the predicted
presence of central density cusps and the overabundance of small
scale structure; see for example \cite{Navarro:1996gj},
\cite{Moore:1999gc} and \cite{Romanowsky:2003qv}. In the light
of these issues, some authors, e.g. \cite{Hu:2000ke} have suggested
that the  Dark Matter could instead consist of ultralight particles
possessing a de Broglie wavelength sufficiently large that
quantum-mechanical effects might manifest themselves on
astrophysically interesting scales. Such models would naturally
predict smoother and less centrally concentrated galaxy haloes than in the CDM case.

Advocating a particular version of this idea,
\cite{Silverman:2002qx} suggested a symmetry-breaking mechanism for
the production of such a particle, based upon a real-valued scalar
field. Although in this case the symmetry-breaking mechanism
provides a nice example of particle production in a universe with a
cosmological constant, symmetry-breaking with a real scalar field
generically produces a catastrophic domain wall problem as argued by \cite{Vilenkin:1986hg}, and this example would seem to be no
exception so this is probably not a viable scenario; see \cite{Brook:2008PhD} for a further discussion. However, these papers consider the possibility that the
Dark Matter component resides in a Bose-Einstein condensate (BEC).
The dynamics and possible observational consequences of a
cosmological fluid with such properties has been investigated
by \cite{Boehmer:2007um}, using techniques developed in the field of
 condensed matter physics. The equation describing a BEC is known to
 condensed matter theorists as the Gross-Pitaevskii (GP) equation, but is probably
more familiar to cosmologists as the nonlinear Schr{\"o}dinger
equation (NLSE).

In condensed matter theory, the term Bose-Einstein condensate  is
usually applied to a dilute bosonic gas confined by an external
potential, the  bosons occupying the lowest available quantum state.
Typically, in the limit of large particle number, the density
distribution of the condensate is taken to be described by a
macroscopic wave-function that is considered to be a quantum field.
This field is manipulated by the Gross-Pitaevskii equation, or
nonlinear Schr{\"o}dinger equation, rather than working with the
usual creation and annihilation operators of quantum mechanics. The
density distribution of the condensate can be represented by a
macroscopic wave-function of the same form as the ground state
wave-function of a single particle. The momentum distribution of the
condensate is obtained by taking the Fourier transform of this
wave-function. In an experimental setup, the occurrence of a
Bose-Einstein condensate is confirmed by a sharp peak in the
momentum space distribution of the gas of particles.

More speculatively, the concept of a BEC can also be applied to such
hypothetical particles  as axions or ghosts. In this context, the
axion field, for example, is coherent and has relatively small
spatial gradients. The gradient energy can be interpreted as
particle momenta, which will be the same and small for each
particle, hence giving a sharp peak in the momentum space
distribution as in the case of the more familiar BEC described
above.

In  quantum field theory, a condensate corresponds to a non-zero
expectation value for some operator in the vacuum and, in the limit
of large occupation number, this condensate can be considered to be a
classical field. This is a good model for the condensate of Cooper
pairs in a superconductor, or for helium atoms in a superfluid, as shown by
\cite{Pethick:2008}.

The usual linear Schr{\"o}dinger equation, coupled to the Poisson
equation, can be used to model many phenomena in cosmology. As well
as modelling a quantum mechanical system, as in \cite{Hu:2000ke}, it
has also been used as a classical wave equation to model structure
formation. It has been shown that using the condensed matter concept
of a Madelung transformation to yield the Euler and continuity
equations from the Schr{\"o}dinger equation, applies as well as to a
cosmological fluid as it does to fluids in the laboratory. For further discussion see
\cite{Spiegel:1980fd,Widrow:1993qq,Coles:2001fw,Coles:2002as,Coles:2002sj,Short:2006md}.

\cite{Silverman:2002qx} also considered the rotation
of a galactic-scale dark matter halo. Using a phenomenological
description taken directly from condensed matter, they concluded
that a galactic halo should be threaded by a lattice of quantised
vortices as a consequence of the rotation of that galaxy. Indeed,
from studies of rotating BECs and quantum turbulence by
\cite{Vinen:2000ts} and  \cite{Short:2007PhD}, it would seem to be difficult to
prevent such vortices from forming. The galaxy velocity rotation
curve produced by these authors reproduces the approximate form of
observed rotation curves.

A similar conclusion was reached in \cite{Yu:2002sz}.
This paper considered stationary cylindrical solutions of a
complex $\phi^{4}$ scalar field model, coupled to gravity.
These solutions are Nielson-Olesen vortices, also known as local U(1)
Cosmic Strings; see \cite{Vilenkin:1986hg}. To describe the motion of these vortices
in the galaxy, Yu and Morgan's procedure was to calculate the motion of
one vortex according to a gradient in the phase induced by the surrounding vortices.

There are many models using the Schr{\"o}dinger-Poisson, or the
relativistic Einstein-Klein-Gordon system to describe slightly
different physical processes. A non-exhaustive list includes scalar
field dark matter (\cite{Matos:1998vk,Lesgourgues:2002,Guzman:2003kt}), boson stars
(\cite{Seidel:1990jh}), Oscillatons (\cite{Seidel:1991zh}); condensate
stars (\cite{Mazur:2001fv}), repulsive dark matter
(\cite{Goodman:2000tg}) and fluid dark matter
(\cite{Arbey:2003sj,Peebles:2000yy}), as well as the fuzzy dark matter
and classical fluid approaches that we have already mentioned, and
the more established theories such as the Abelian Higgs model in
field theory, and the Landau-Ginzberg model in condensed matter. We
will not attempt a thorough review of each model here, except to say
that it is sometimes difficult to explicitly distinguish between
them.

The effects of the interaction of gravity with a coherent state of
matter, such as a BEC, have certainly been considered
(\cite{Carroll:PC,Carroll:CV}), and prompted the question of whether
it is actually possible for DM to be in a coherent quantum state if
the only interaction with visible matter is gravitational. Penrose
has also used the Schr{\"o}dinger-Poisson system during his `Quantum
State reduction' research program; see \cite{Moroz:1998dh}.

In this paper we seek to determine some of the properties of a
quantised vortex residing in a galactic-scale Bose-Einstein
condensate dark matter. In particular, we will place bounds on the
parameters that are used to describe such a vortex. For the purposes
of this paper we presume that the DM does indeed consist of a BEC,
formed at an earlier stage of cosmological history and described by
the coupled nonlinear Schr{\"o}dinger-Poisson system, and that
vortices are present in this cosmological fluid.

In Section \ref{Setup} we introduce the basic formalism for
describing a BEC using the Gross-Pitaevskii (nonlinear
Schr{\"o}dinger) equation, and vortices within it. In Section
\ref{Gravitationally Coupled BECs} we discuss coupling the NLSE to
the Poisson equation. In Sections \ref{Vortex Stability in
Gravitationally Coupled BECs} and \ref{Bounds on Parameters} we
look at some of the properties of a vortex as a result of
gravitational coupling. We present some results in Section
\ref{Results} and a discussion in Section \ref{Discussion}, including updates
from the literature since the original version of this paper was published.

\section{Setup}
\label{Setup}

For our discussion, we use some of the conventions and procedures set out by \cite{RobertsBerloff} and \cite{Pethick:2008}. The nonlinear Schr{\"o}dinger
equation is written in the form
\begin{equation}
i \hbar \Psi_{t}=-\frac{\hbar^{2}}{2m} \nabla^{2} \Psi + \Psi \int
|\Psi(x',t)|^{2} V(|x-x'|)dx',
\end{equation}
where $m$ is the mass of a particle in the BEC, and $V(|x-x'|)$ is the
interaction potential between bosons. The potential is simplified
for a weakly interacting Bose system by replacing $V(|x-x'|)$ with a
$\delta$-function repulsive potential of strength $V_0$, giving
\begin{equation}
i \hbar \Psi_{t}=-\frac{\hbar^{2}}{2m} \nabla^{2} \Psi + V_0
|\Psi|^{2}\Psi.
\end{equation}
Defining a state that is independent of time to be the `laboratory
frame', $\Psi = \exp(iE_{\upsilon}/\hbar)$, it is then possible to
consider deviations from that state by considering the evolution of
$\psi$, where $\psi = \Psi \exp(iE_{\upsilon}t/\hbar)$. Here,
$E_{\upsilon}$ is the chemical potential of a boson, in the sense
that it is the increase in ground state energy when one boson is
added to the system. The nonlinear Schr{\"o}dinger equation used for
subsequent analysis is then
\begin{equation}\label{NLSe}
i \hbar \psi_{t}=-\frac{\hbar^{2}}{2m} \nabla^{2} \psi
+V_{0}|\psi|^{2}\psi - E_{\upsilon}\psi.
\end{equation}

Multiplying equation (\ref{NLSe}) by $\psi^{*}$ and subtracting the complex conjugate of the resulting equation we obtain
\begin{equation}
\frac{\partial |\psi|^{2}}{\partial t} = {\bf \nabla} . \left[ \frac{\hbar}{2mi}(\psi^{*} {\bf \nabla} \psi - \psi {\bf \nabla}{\psi^{*}}) \right].
\end{equation}
We notice that this is of the form of a continuity equation.
\begin{equation}
\frac{\partial |\psi|^{2}}{\partial t} + {\bf \nabla}. (|\psi|^{2} {\bf v}).
\end{equation}
We identify $|\psi|^{2}$ as the number density $n$, and the related momentum density is given by
\begin{equation}\label{flux}
{\bf j} = \frac{\hbar}{2i}(\psi^{*} {\bf \nabla} \psi - \psi {\bf \nabla}{\psi^{*}}),
\end{equation}
which is equivalent to
\begin{equation}\label{fluxvel}
{\bf j} = mn {\bf v}.
\end{equation}
This defines for us the mass density as $\rho = m n = m |\psi|^{2}$, and the velocity
\begin{equation}\label{veldef}
{\bf v} = \frac{\hbar}{2mi} \frac{(\psi^{*} {\bf \nabla} \psi - \psi {\bf \nabla}{\psi^{*}})}{|\psi|^{2}}.
\end{equation}
As suggested in the Introduction, we can make a `Madelung transformation'
\begin{equation}
\psi = \alpha \exp \left(i \phi_{\omega}\right),
\end{equation}
and from equation (\ref{veldef}), we obtain an expression for the velocity of the condensate
\begin{equation}
{\bf v} = \frac{\hbar}{m} {\bf \nabla}\phi_{\omega}.
\end{equation}
Here, $\phi_{\omega}$ is the velocity potential. Substituting the Madelung transformation and taking real and imaginary parts yields the fluid equations: the continuity equation
\begin{equation}
\frac{\partial \left(\alpha ^{2}\right)}{\partial t} + \frac{\hbar}{m} \nabla . (\alpha^{2}
\nabla \phi_{\omega})=0;
\end{equation}
and the (integrated) Euler equation:
\begin{equation}
\hbar \frac{\partial \phi_{\omega}}{\partial t} = \frac{\hbar^{2}}{2m}\frac{\nabla^{2} \alpha}{\alpha}-\frac{1}{2}m {\bf v}^{2} - V_{0}\alpha^{2} + E_{\upsilon}.
\end{equation}
Often the identification
\begin{equation}
{\phi_{\omega}}' = \frac{\hbar}{m} \phi_{\omega}
\end{equation}
is used, to maintain contact with the more familiar form of the fluid equations:
\begin{equation}
\frac{\partial \left( \alpha^{2} \right)}{\partial t}+ \nabla . (\alpha^{2}
\nabla {\phi_{\omega}}')=0,
\end{equation}
\begin{equation}
\frac{\partial {\phi_{\omega}}'}{\partial t} =
\frac{\hbar^{2}}{2m^2}\frac{\nabla^{2} \alpha}{\alpha}-\frac{(\nabla
{\phi_{\omega}}')^{2}}{2}-\frac{V_{0}}{m} \alpha^{2}+\frac{E_{\upsilon}}{m}.
\end{equation}
Here the quantum nature of the fluid is evident only in the first term on the right hand side of the second equation, which is often known as the {\em quantum pressure} term, although dimensionally it is a chemical potential. This term is relevant only on small scales, where quantum effects become important, such as in a vortex core, or where the condensate meets a boundary. This identification rather hides the quantum nature of the fluid with respect to the fluid velocity, which will become particularly relevant when we start talking about vortices in the next section.

By assuming that the condensate reaches a
stationary equilibrium state at a distance far from any disturbance,
equation (\ref{NLSe}) gives us the relation
\begin{equation}\label{psiinfty}
\psi_{\infty} = \left( \frac{E_{\upsilon}}{V_{0}} \right)^{\frac{1}{2}}.
\end{equation}
When the condensate wave-function reaches a boundary, such as the
wall of a container, or the core of a vortex is being considered, we
can define a distance over which the wave-function changes from zero
to its bulk value, or where quantum effects become important:
\begin{equation}\label{healinglength}
a_{0}=\frac{\hbar}{(2mE_{\upsilon})^{\frac{1}{2}}}
\end{equation}
This is known as the {\em coherence length}, or {\em healing
length}, as it is the distance over which the wave-function requires
`healing';  see \cite{RobertsBerloff} and \cite{Pethick:2008}.

\subsection{Vortices}

We have already seen that the velocity of the condensate is given by
\begin{equation}
{\bf v} = \frac{\hbar}{m} {\bf \nabla}\phi_{\omega}.
\end{equation}
One would expect then, that the condensate would be irrotational, as
\begin{equation}
{\bf \nabla} \times ({\bf \nabla} f) = 0
\end{equation}
for any scalar, $f$. This restricts the motion of the condensate much more than a classical fluid. The circulation around any contour then, should also be zero. By Stokes' theorem
\begin{equation}
\Gamma = \oint_{l} {\bf v}. d {\bf l} = \int_{A} ({\bf \nabla} \times {\bf v}). d {\bf A} = 0.
\end{equation}
This condition, known as the Landau state, was first derived in an
analysis of superfluid HeII by \cite{Landau:1941}, and it suggests that
rotation of such a condensate should be impossible. Experiments by
\cite{Osbourne:1950} indicated that the condensate did
indeed experience rotation. Building on the independent work of \cite{Onsager:1949},  \cite{Feynman:1955} suggested that
rotation and hence non-zero circulation could be explained by
assuming that the condensate is threaded by a lattice of parallel
vortex lines. It is possible to have circulation surrounding a
region from which the condensate is excluded and in this case this
would be the vortex core. To see this, we note that the condensate
wave-function must be single-valued, and so around any closed
contour the change in the phase of the wave-function $\Delta \phi$
must be a multiple of 2$\pi$.
\begin{equation}
\Delta \phi_{\omega} = \oint {\bf \nabla}  \phi_{\omega} . d{\bf l} = 2\pi l
\end{equation}
where $l$ is an integer. We immediately see that the circulation is quantised in units of $h/m$.
\begin{equation}
\Gamma = \oint {\bf v}. d{\bf l} = \frac{\hbar}{m} 2 \pi l = l \frac{h}{m}
\end{equation}
To obtain vortex solutions, we work in cylindrical coordinates $(r,
\chi, z)$ and look for a static solution of the nonlinear
Schr{\"o}dinger equation, equation (\ref{NLSe}). To satisfy the
requirement of single-valuedness, the condensate wave-function must
vary as $\exp (i n \chi)$, with $n$ integer. We make the vortex
ansatz
\begin{equation}\label{ansatz}
\psi = R(r) \exp (i n\chi).
\end{equation}
It is interesting to note the similarity between this procedure, and
that used in obtaining Nielson-Olesen vortices, or Cosmic Strings,
in the Abelian-Higgs model discussed by \cite{Vilenkin:1986hg}. This was mentioned in Section \ref{Introduction}, and will be useful shortly for obtaining equation (\ref{densityapprox}), as shown in Section \ref{Approximations to the Density Profile}.
We can obtain an expression for the velocity of a vortex by substituting the vortex ansatz, equation (\ref{ansatz}) into equation (\ref{veldef})
\begin{equation}\label{vV}
{\bf v}_{\omega} =  \frac{\hbar n}{r} \frac{1}{m} \bf{\hat{\chi}},
\end{equation}
and we note again the discrete nature of the allowed values of velocity.
From now on we will consider only $n = 1$ vortices. Vortices with $n > 1$ are generally expected to be unstable, from energy considerations - see for example Chapter 9.2.2 of \cite{Pethick:2008} - and will break up into several  $n = 1$ vortices to make up a vortex lattice, as described above. We can note further that Cosmic Strings with winding numbers $n > 1 $ are also unstable to perturbations \cite{Vilenkin:1986hg}. Such defects break down to several $n = 1$ configurations in both a condensed matter environment, and a high-energy field-theoretic one. Feynman initially introduced quantised vortices as a purely theoretical tool with which to explain the rotation of the condensate, but the experimental verification of the quantisation of rotational velocities (by, e.g. \cite{Packard:1972}, demonstrated that these vortices were indeed real.
The density profile $\rho(r) = m |R(r)|^2$ is defined by the vortex equation, which results from substituting the vortex ansatz into equation (\ref{NLSe})
\begin{equation}\label{densityprofilevortex}
-\frac{\hbar^{2}}{2mE_{\upsilon}} {\Bigg[}\frac{d^{2}R(r)}{dr^{2}}+\frac{1}{r} \frac{dR(r)}{dr}- \frac{1}{r^{2}}R(r) {\Bigg]}+\frac{V_{0}}{E_{\upsilon}}{R(r)}^{3} - R(r) = 0
\end{equation}
From equation (\ref{psiinfty}) we see that the density far from the
vortex is given by
\begin{equation}\rho_{\infty} = m R_\infty = m
\frac{E_{\upsilon}}{V_{0}}.
\end{equation}
Analytic solutions of this equation are not available so it must be solved numerically if very precise results are needed. There are however a number of approximate techniques available.

\subsection{Approximations to the Density Profile}
\label{Approximations to the Density Profile}

The numerical solution to the NLSE can be cumbersome to work with,
so we provide some discussion of some approximations that can be
used. It is possible to scale the the variables $r$ and $R(r)$ in
equation (\ref{densityprofilevortex}) to obtain a scale-free
equation. Scaling $r$ by the healing length, $r' = r/a_{0}$, and
$R(r)$ by the steady state value, $R'(r') = R(r)/R_{\infty}$ we
obtain
\begin{equation}\label{densityprofilevortexscaled}
\frac{d^{2}R'(r')}{{dr}'^{2}}+\frac{1}{r'} \frac{dR'(r')}{dr'}-
\frac{R'(r')}{{r'}^{2}} - {R'(r')}^{3} + R'(r') = 0.
\end{equation}
Our first idea for an approximation comes from the field of cosmic
strings. The method of approximation is detailed in
\cite{Vilenkin:1986hg}. Looking at the profile of the Higgs field in
a Nielson-Olesen vortex we see that it can be written, in a
similarly scaled way, as
\begin{equation}\label{VSCSScaled}
\frac{d^{2}R'(r')}{dr'^{2}}+\frac{1}{r'} \frac{dR'(r')}{dr'}- \frac{R'(r')}{r'^{2}} (A(r')-1)^{2} - \frac{\lambda}{2}R'(r')({R'(r')}^{2} -1) = 0
\end{equation}
Here $A(r')$ is a gauge term arising from the coupling to
electromagnetism, and $\lambda$ is determined by the potential term
of the theory. It is possible to linearise equation
(\ref{VSCSScaled}) to obtain a modified Bessel function as the first
order approximation to R'(r') - the zeroth order being 1. This
happens in the string case, because the gauge contributions serve to
cancel one of the terms, leaving the modified Bessel's equation. The
linearised version of equation (\ref{densityprofilevortexscaled})
does not quite reduce to a modified Bessel's equation, but taking
our lead from the cosmic string example, we write
\begin{equation}\label{ScaledApproximation}
R'(r') \sim 1 - \exp (-r').
\end{equation}

\begin{figure}[ht]
\begin{center}
\includegraphics[width=1\textwidth]{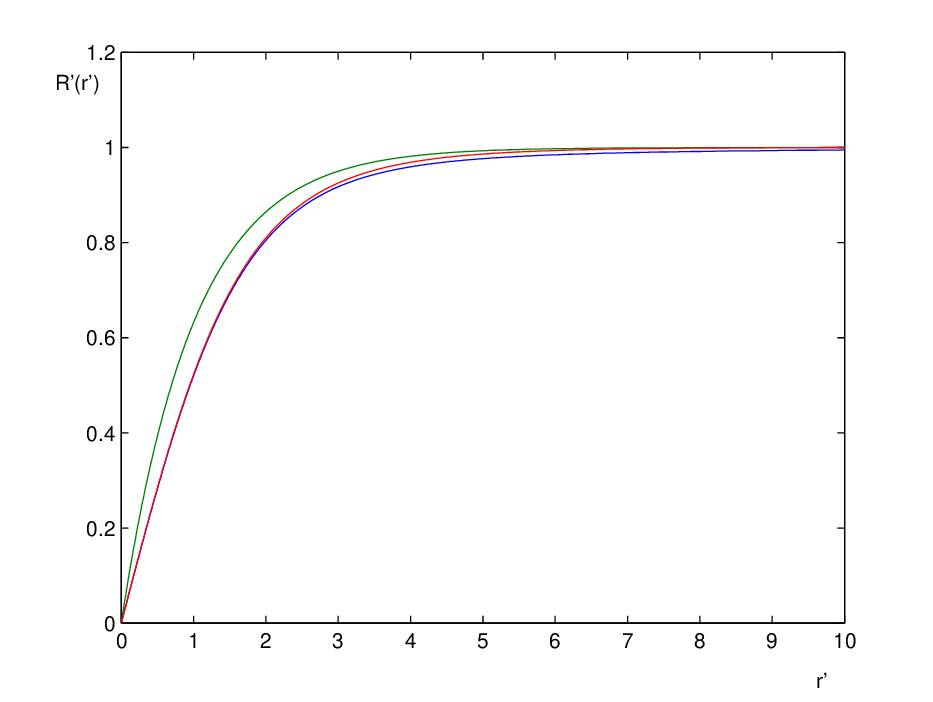}
\caption{Numeric solution to equation (\ref{densityprofilevortexscaled}) (blue), the Pad{\`e} approximation equation (\ref{Pade}) (red), and the scaled approximation used in this analysis, equation (\ref{ScaledApproximation}) (green).}
\label{ScaledDensityProfilefig}
\end{center}
\end{figure}

Another approximation, which might seem to be more accurate, was
developed by \cite{BerloffPade} in a condensed matter
context. The Pad{\'e} approximation has the same asymptotics at $r =
0$ and $r = \infty$ as the function one is trying to approximate.
The Pad{\'e} approximation in this case gives
\begin{equation}\label{Pade}
R'(r') \sim \sqrt{\frac{{r'}^2 (0.3437 + 0.0286 {r'}^2)}{1+ 0.3333
{r'}^2 + 0.0286{r'}^4}}.
\end{equation}
This solution is plotted in Figure \ref{ScaledDensityProfilefig} along with the numeric solution given by equation (\ref{densityprofilevortexscaled}), and the previous approximation, equation (\ref{ScaledApproximation}).
The Pad{\'e} approximation is indeed much more accurate in
the small and large $r$ regions.
However, the Pad{\'e} approximation has the tendency to overestimate
the density in the central region, producing a density function whose
derivative is negative in this region.
As discussed in the main body of this paper, the gravitational potential
is proportional to the density, and so the gravitational force will be
proportional to the derivative of the density function.
If we chose to use the Pad{\'e} approximation for our density profile,
we could be potentially misled by its behaviour in the central region.

To illustrate the results obtained in the following, we will therefore use the approximation
\begin{equation}
R(r) = \left(\frac{E_{\upsilon}}{V_{0}}\right)^{\frac{1}{2}} \left[1 -
\exp[-r/a_{0}]\right),
\label{densityapprox}
\end{equation}
though the bounds we obtain are relatively insensitive to this choice.

\section{Self-gravity of a BEC Vortex}
\label{Gravitationally Coupled BECs}

In considering Bose-Einstein condensates on scales relevant to
structure formation in the universe, we must necessarily include
gravitational effects. BECs are typically sufficiently dilute that
the mass densities are not very large, and so a Newtonian
approximation is sufficient. Gravitational effects can be added to
the BEC by including a term in the nonlinear Schr{\"o}dinger
equation that couples to the Poisson equation. We then have a pair
of equations modelling a gravitationally coupled fluid.
\begin{equation}\label{NLSP1}
i \hbar \psi_{t}=-\frac{\hbar^{2}}{2m} \nabla^{2} \psi +V_{0}|\psi|^{2}\psi - E_{\upsilon}\psi + m\phi_G \psi
\end{equation}
\begin{equation}\label{NLSP2}
\nabla^2 \phi_G = 4 \pi G \rho = 4 \pi G m |\psi|^{2}.
\end{equation}

\subsection{Vortices in Gravitationally Coupled BECs}
\label{Vortices in Gravitationally Coupled BECs}

To obtain vortex solutions, we again work in cylindrical coordinates
$(r, \chi, z)$, and substitute the vortex ansatz $\psi = R(r) exp (i
\chi)$ into equations (\ref{NLSP1}) and (\ref{NLSP2}). The system of
equations describing a gravitationally coupled BEC fluid becomes
\begin{equation}\label{NLSPV1}
-\frac{\hbar^{2}}{2mE_{\upsilon}} {\Bigg[}\frac{d^{2}R(r)}{dr^{2}}+\frac{1}{r} \frac{dR(r)}{dr}- \frac{1}{r^{2}}R(r) {\Bigg]}+\frac{V_{0}}{E_{\upsilon}}{R(r)}^{3} - R(r) + m \phi_{G} (r)=0
\end{equation}
\begin{equation}\label{NLSPV2}
\nabla^{2} \phi_{G} (r) = \frac{d^{2} \phi_{G}(r)}{dr^{2}} + \frac{1}{r} \frac{d \phi_{G (r)}}{dr} = 4 \pi G m R(r)^{2}
\end{equation}
Ideally, we would like to find a solution describing the function
$R(r)$ in this system, so we can compare the density profile of a
quantum vortex, to that of one that is gravitationally coupled.
However, finding a full simultaneous solution to these coupled
equations is difficult. Firstly, because the nonlinear
Schr{\"o}dinger equation itself is not soluble analytically.
Secondly, because the vortex density tends to a constant,  so the
Newtonian potential tends to diverge, and thirdly because these
equations do not define the vortex velocity, which would be
providing the centripetal force to withstand the gravitational
collapse. In other words, all the variables required to provide a
fully simultaneous static solution are not defined within these two
equations.

\section{Vortex Stability in Gravitationally Coupled BECs}
\label{Vortex Stability in Gravitationally Coupled BECs}

Rather than solving the coupled equations (\ref{NLSP1}) and
(\ref{NLSP2}) directly, we can make some arguments regarding the
stability of a gravitationally coupled BEC vortex, and consequently
give some bounds on the parameters that describe it. Our analysis is
based upon consideration of the radial velocity profile of a BEC vortex,
$v_{\omega}(r)$, and the radial velocity induced from gravitational
attraction, $v_{\rm G}(r)$; $v_{\omega}(r)$ is the velocity that
the vortex density distribution is moving at, for a particular $r$,
while $v_{\rm G}(r)$ would be the velocity experienced by a test
particle orbiting that density distribution, at a distance $r$. To
sustain a vortex, $v_{\omega}(r)$ must at least be greater than
$v_{\rm G}(r)$, otherwise the quantum-mechanical forces at work in
the vortex are not sufficiently strong to hold itself up against
gravitational collapse. That is, the vortex is spinning too slowly
to provide enough centripetal force to balance the gravitational
force. For stability, we therefore have the bound,
\begin{equation}\label{velocityequality}
v_{\omega}(r) \geq v_{\rm G}(r)
\end{equation}

\subsection{Gravitational Field of a Cylindrically Symmetric System}

To obtain $v_{\rm G}(r)$, we turn to Gauss's Law to determine the
gravitational field of a cylindrically symmetric mass distribution,
and hence obtain the radial gravitational velocity of a test
particle moving in the field of that system. Gauss's law is
\begin{equation}\label{GausssLaw}
\oint {\bf g} \cdot d{\bf A} = -4\pi GM_{\rm encl}
\end{equation}
The density, $\rho(r) = m |R(r)|^{2}$, is already
determined in terms of the cylindrical r co-ordinate, as it is a
solution of the vortex equation. The mass enclosed is the
density pervading a cylinder of radius r and length L.
\begin{equation}
M_{\rm encl}= L \int_{0}^{r} 2 \pi r \rho (r) dr
\end{equation}
The left-hand side of Gauss's law, in cylindrical co-ordinates, is
\begin{equation}
\int g r d \chi dz,
\end{equation}
where the integral over the z co-ordinate is again L, the length of the vortex.
Gauss's law, then, gives us
\begin{equation}
g r 2 \pi L = -4 \pi G 2 \pi L \int_{0}^{r}\rho(r) r dr
\end{equation}
giving
\begin{equation}
g= - \frac{4 \pi G m }{r} \int_{0}^{r}|R(r)|^{2} r dr.
\end{equation}
The sign is negative as we have chosen an outward-pointing surface
normal in our formulation of Gauss's Law, equation
(\ref{GausssLaw}), which indicates that the gravitational flux will
always be towards the origin. This leads to the slightly
counter-intuitive conclusion that a hole (the vortex) in a constant
mass density background would seem to produce a gravitational force
towards it, but this is really a manifestation of the (extremely)
thick shell condition. Viewed another way, this static configuration
will want to act to collapse in, and close the hole. It is this
force that is `unopposed' in equations (\ref{NLSPV1}) and
(\ref{NLSPV2}). This need not concern us further, as it is the magnitude that is required for our argument. The magnitude of the induced
centripetal force is \begin{equation} g=\frac{{v_{\rm G}}^{2}}{r},
\end{equation}
and the gravitational circular velocity profile $v_{\rm G}$ is given
by
\begin{equation}\label{vG}
{{v_G}(r)}^2 = 4 \pi G \int_{0}^{r} \rho(r) r dr = 4 \pi G m
\int_{0}^{r} |R(r)|^{2} r dr.
\end{equation}

\section{Bounds on Parameters}
\label{Bounds on Parameters}

We now have expressions for $v_{\rm G}(r)$ and $v_{\omega}(r)$,
equations (\ref{vV}) and (\ref{vG}), to go in the bound given by
equation (\ref{velocityequality}). 

To enable us to obtain actual values for the velocity and density
profiles that we are considering, we must provide values for the
parameters $m$, $V_{0}$, and $E_{\upsilon}$. The properties of Dark Matter
particles are, by their very nature, unknown, so we must make some
approximations. We use the analysis in \cite{Silverman:2002qx} to
provide us with some data values. The mass of the Bose Einstein
condensate Dark Matter particle in that paper is 3.56 $\times$
10$^{-59}$ kg (2 $\times$ 10$^{-23}$ eV). Their analysis is based on the mass and angular
rotation of the Andromeda galaxy. The mean density is given as 2
$\times$ 10$^{-24}$kg m$^{-3}$, and they estimate that the vortex
line density in the galaxy would be about 1 vortex per 208 kpc$^2$.
This gives a vortex radius of  $r_{\omega} \sim$ 2.5 $\times$
10$^{20}$ m. We again turn to vortex lattices in condensed matter
systems to provide us with some further estimates of vortex
properties in a BEC.

Taking the distance between two vortices to be twice the vortex
radius, we note from experimental observations of vortex lattices in
a BEC that the vortex density reaches the normal density at about
half the vortex radius; see, for example, Figure 9.3 in
\cite{Pethick:2008}, taken from \cite{Coddington:2003}.  From Figure
(\ref{ScaledDensityProfilefig}), we also see that the vortex density
reaches the normal condensate density at around five healing
lengths. This gives us an estimate of ${r_{\omega}}/{2} =
5a_{0}$. We then use   $r_{\omega}\sim$ 2.5 $\times$ 10$^{20}$ m,
$a_{0}={\hbar}/(2mE_{\upsilon})^{\frac{1}{2}}$, and
$\rho_{\infty} = m {E_{\upsilon}}/{V_{0}}$ to give estimates for
$E_{\upsilon}$ and $V_{0}$. With these approximations we find values of
$E_{\upsilon}$ $=$ 2.5 $\times$ 10$^{-49}$ J (1.56 $\times$ 10$^{-30}$eV) and $V_{0} = 4.45
\times$ 10$^{-84}$ J m$^3$ (3.7 $\times$ 10$^{-45}$ eV$^{-2}$).

\begin{figure}[ht]
\begin{center}
\includegraphics[width=1\textwidth]{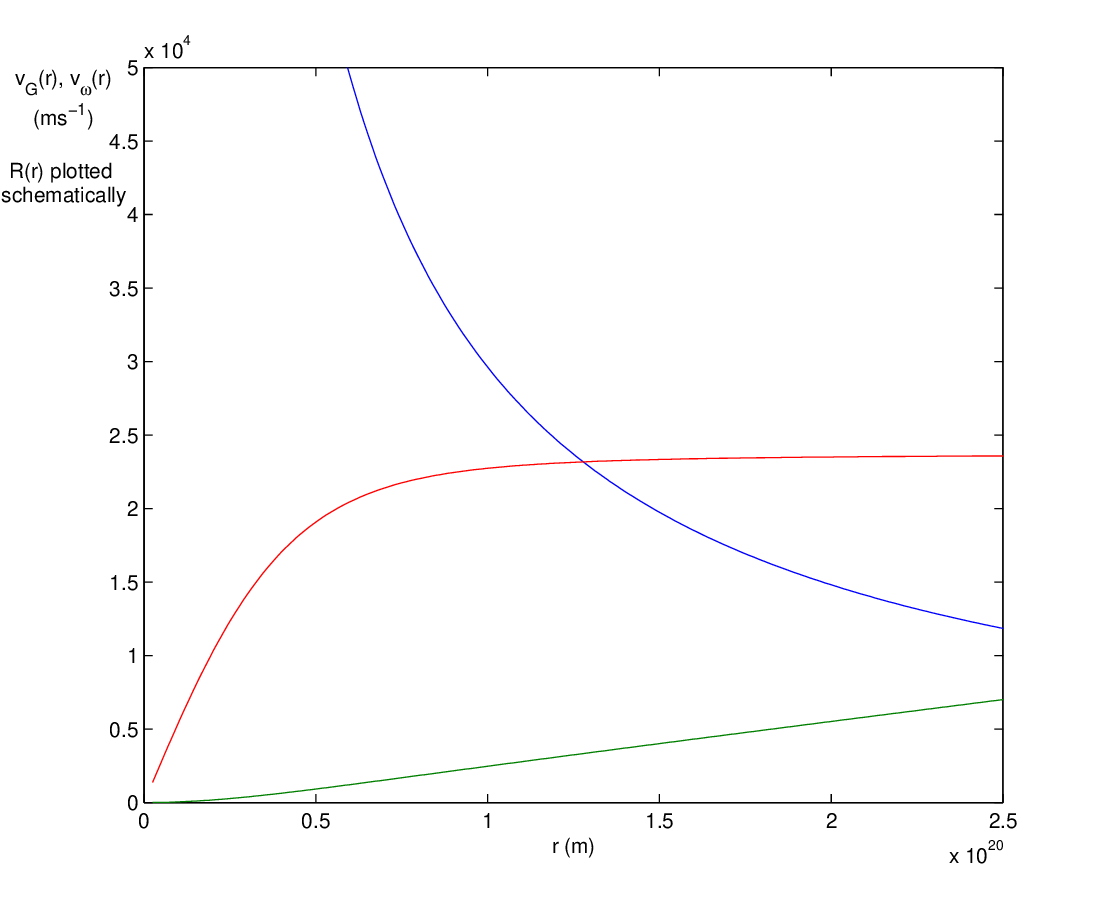}
\caption{Velocity Profiles for v$_{G}$ (green) and $v_{\omega}$
(blue). Density profile plotted schematically for comparison (red).}
\label{VortexVelProfiles}
\end{center}
\end{figure}

In Figure \ref{VortexVelProfiles}
we plot, as an example, $v_{\omega}(r)$ and $v_{\rm G}(r)$ and the
density profile for comparison. For this example, we have used
values of m $=$ 3.56 $\times$ 10$^{-59}$kg (2 $\times$ 10$^{-23}$ eV),  E$_{\upsilon}$ $=$ 2.5
$\times$ 10$^{-49}$ J (1.56 $\times$ 10$^{-30}$eV) and $V_{0}$ $=$ 4.45 $\times$ 10$^{-84}$ Jm$^{3}$ (3.7 $\times$ 10$^{-45}$ eV$^{-2}$)
as explained above.

The bound on stability, $v_{\omega}(r) \geq v_{\rm G}(r)$, will
always be violated at some point, as outside the vortex core
$v_{\omega}(r) \sim 1/r$ and  $v_{\rm G}(r) \sim r$. We must
specify what might be an acceptable value of r for  $v_{\omega}(r)$
and $v_{\rm G}(r)$ to meet. For a vortex to exist, the density
profile should be fully established. We take this to mean that the
density has essentially reached its background level. From the
scaled density profile in discussed in \ref{Approximations to the
Density Profile}, and plotted in Figure
(\ref{ScaledDensityProfilefig}), we see that the density reaches its
background level at a value of about ten times the healing length.
Using equation (\ref{densityapprox}) in (\ref{vG}), equations
(\ref{vG}) and (\ref{vV}) in (\ref{velocityequality}), and substituting for $E_{\upsilon}$ from equation (\ref{healinglength})  we obtain
\begin{equation}\label{plottingVm}
\frac{\sqrt{2\pi}}{2} \left(\frac{G \hbar^2}{V_{0}{a_{0}}^2} \left[2r^2
+ 8r{a_{0}}e^{-\frac{r}{{a_{0}}}} + 8{a_{0}}^{2}
e^{-\frac{r}{{a_{0}}}} - 2r{a_{0}} e^{-\frac{2r}{{a_{0}}}} -
{a_{0}}^{2} e^{-\frac{2r}{{a_{0}}}}\right] \right)^{\frac{1}{2}}
\leq \frac{\hbar}{mr}.
\end{equation}
We will fix the healing length $a_0$, and plot $V_{0}$ against $m$
(fixing $a_{0}$ and $m$ fixes $E_{\upsilon}$, from equation
(\ref{healinglength})) to give an allowed range of parameter
values. We will do this for various values of $a_0$, and for various
values of $r$, which we will take to be an integer number of healing
lengths, $r=na_0$, with the minimum $n = 10$ as outlined above.
Equation (\ref{plottingVm}) then becomes
\begin{equation}\label{Vleq}
V_{0} \geq \frac{\pi}{2} G m^2 n^2 \left(2n^2 {a_{0}}^2 + 8n {a_{0}}^2
e^{-n} + 8{a_{0}}^2 e^{-n} - 2n {a_{0}}^2 e^{-2n} -  {a_{0}}^2
e^{-2n}\right).
\end{equation}

\begin{figure}[ht]
\begin{center}
\includegraphics[width=1\textwidth]{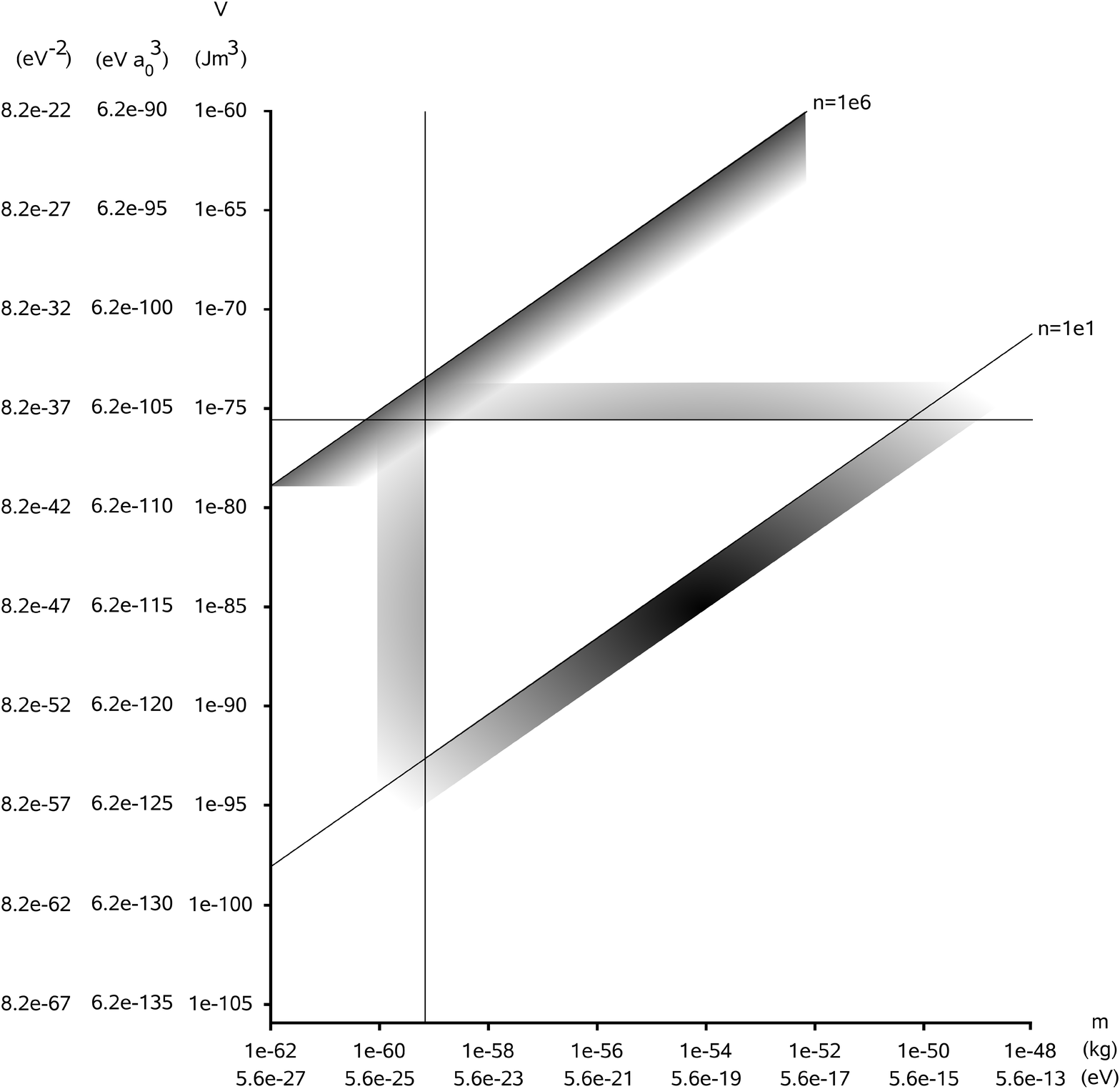}
\caption{Allowed region in $V_{0} - m$ parameter space, for a healing length
of $a_{0} = 1 \times 10^{16}$ m ($\sim$ 1 parsec) } \label{Vma3real}
\end{center}
\end{figure}

\subsection{Other Bounds}
We can obtain some other bounds to cut off other bits of parameter space.
The asymptotic vortex density is given by
\begin{equation}\label{backgrounddensity}
\rho_{\infty} = m\left(\frac{E_{\upsilon}}{V_{0}}\right).
\end{equation}
If the vortex exists as a component of a galaxy, then there is a
minimum and maximum density that the vortex can have, given by the
maximum and minimum known values of mass density within the system:
\begin{equation}
\rho_{\rm min} \leq \rho_{\infty} \leq \rho_{\rm max}.
\end{equation}
The value of $E_{\upsilon}$ in equation (\ref{backgrounddensity}) is
fixed (as we are fixing the healing length), and so the bound on the
density becomes a bound on $V_{0}$.
\begin{equation}\label{densitybound}
\frac{\hbar^2}{2 {a_{0}}^{2} \rho_{max} } \leq V_{0} \leq
\frac{\hbar^2}{2 {a_{0}}^{2} \rho_{min} }.
\end{equation}
Equation (\ref{Vleq}) gives a lower bound on $V_{0}$, so to obtain an upper bound,
we use the second half of the above relation.
\begin{equation}\label{Vupper}
V_{0} \leq \frac{\hbar^2}{2 {a_{0}}^{2} \rho_{min} }.
\end{equation}
Another bound is provided because the vortex velocity should never
exceed the speed of light,
\begin{equation}
v_{\omega} = \frac{\hbar}{mr} \leq c.
\end{equation}
This is an extremely conservative bound because the assumption of Newtonian dynamics itself would break down here. A tighter limit would be obtained by using, e.g. the escape velocity from a galactic halo, but this would depend on galaxy mass so for simplicity we stick with the safer bound.

It can be seen from equation (\ref{vV}) that the vortex velocity increases with decreasing radius. This relation breaks down within the vortex core, $a_{0}$, where the vortex velocity diverges. Finding an appropriate description is a topic of some interest in condensed matter theory; see for example \cite{Sadd:1997sv}. We evaluate the maximum vortex velocity at a distance of $5a_{0}$ from the origin. i.e. in a regime where we are sure the relation holds. This gives a bound on the mass.
\begin{equation}\label{mupper}
m \geq \frac{\hbar}{5 c  a_{0}}.
\end{equation}
\subsection{Values}
To see how the restriction on $m$ and $V_{0}$ varies, we can think of a
range of healing lengths that cover all possible scales in a galaxy.
\begin{eqnarray}
1 \times 10^{10} {\rm m} \quad (3.2 \times 10^{-10} {\rm kpc}, \quad \sim 7 \times 10^{-2} {\rm AU})  \leq  a_{0} \\
a_{0}\leq 1 \times 10^{22} {\rm m} \quad (324 {\rm kpc})
\end{eqnarray}
This range of scales takes us from sub solar system, to that of the
largest known galaxies, e.g. IC 1101 in the Abell 2029 cluster discussed by \cite{Uson:1990}. At fixed $a_{0}$ we will also cover a large range of $n$; the number of
healing lengths where the velocity profiles cross. For the bound
given in equation (\ref{Vupper}), we take the minimum density found
within a galaxy to be the cosmological density. This minimum must
necessarily be close to the critical density of the universe.
\begin{equation}
\rho_{\rm min} = \rho_{c} \Omega_c = \frac{3 \Omega_c H_{0}^{2}}{8 \pi G}.
\end{equation}
With $H_{0} = 70$ km s$^{-1}$ Mpc$^{-1}$, and $\Omega_c=1$ this gives a value of
$\rho_{\rm min} = 9.2 \times 10^{-27}$ kg m$^{-3}$. This is a rather conservative bound and is chosen to be applicable to as general a situation possible: one could obtain tighter limits for specific systems by fixing the density by more detailed dynamical considerations; see, e.g. \cite{JohnKam}.

\begin{figure}[ht]
\begin{center}
\includegraphics[width=1\textwidth]{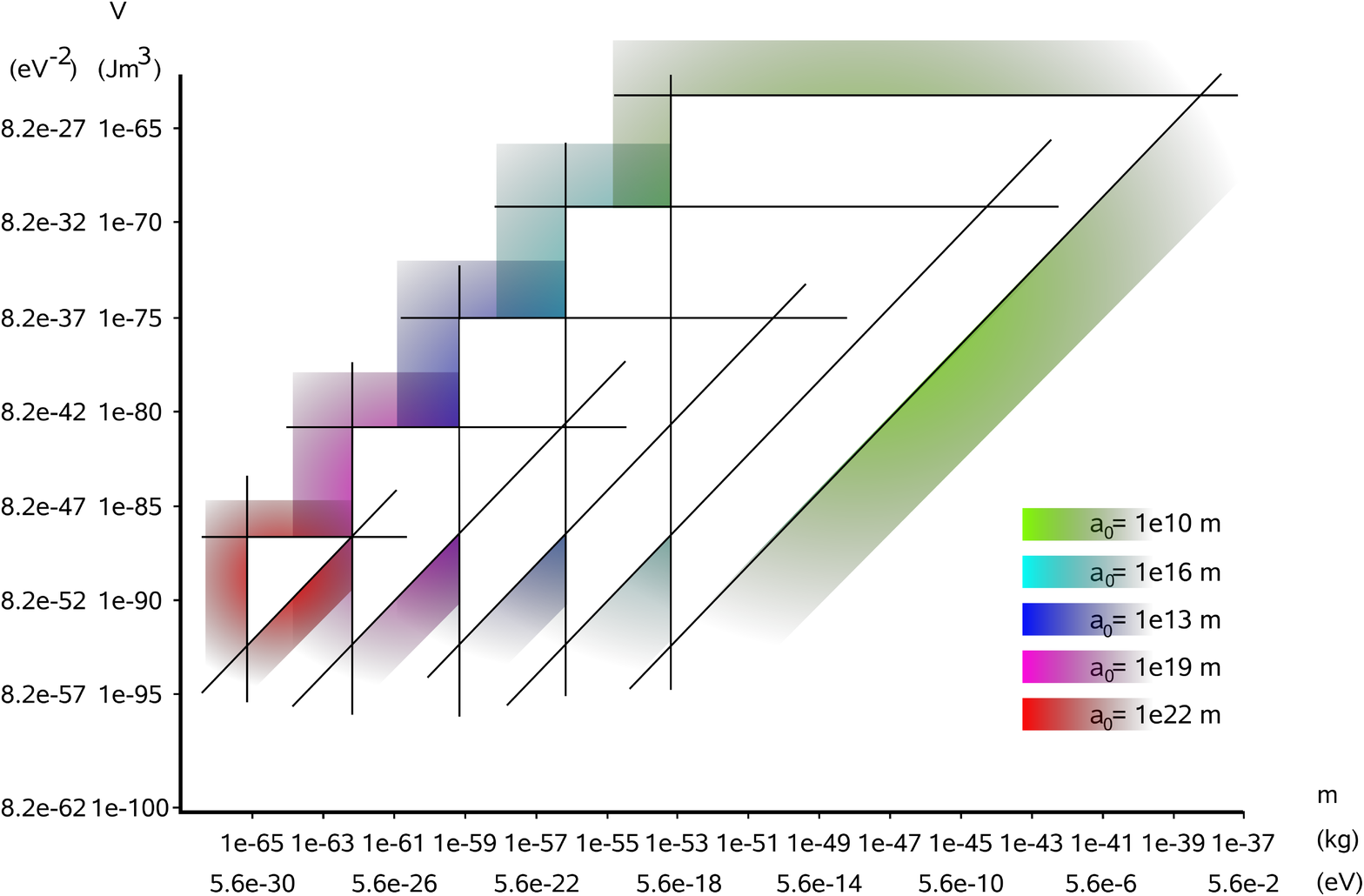}
\caption{Allowed regions in $V_{0}-m$ parameter space, with $n = 10$.
Healing lengths as labelled.} \label{Vmallreal}
\end{center}
\end{figure}

\section{Results}
\label{Results}

The vortex systems we have discussed so far are characterized by a number of parameters that can be related by physical considerations. To simplify the presentation in this paper we choose to take $a_0$ as a free parameter and then use the preceding section to impose bounds on the other parameters. Alternative approaches are possible, including fixing all parameters to be consistent with the size and rotation velocity of galaxies, which would give tighter bounds. For modelling a specific system, such as galaxy like the Milky Way, the range of allowed parameters would be smaller still. One most therefore take the following results to be conservative and illustrative only.

In Figure (\ref{Vma3real}), we show a region of the $V_{0}-m$ parameter
space for the healing length $a_{0} = 1 \times 10^{16}$ m ($\sim$ 1
pc). The lower bound on $V_{0}$ is given when $v_{\omega}$ and $v_G$
cross at a value of ten times the healing length, $n = 10$. A vortex could be
considered more stable if $v_{\omega}$ and $v_G$ cross at a greater
number of n, moving us up into the allowed triangular region.
However, this can soon reach the minimum density bound on $V_{0}$. A value
of $n=10^{6}$ is also plotted, and it is clear that this is outside
the bounded region. The lines bounding the region of allowed
parameter values are given by equations (\ref{Vleq}), (\ref{mupper})
and (\ref{Vupper}).

Figure (\ref{Vmallreal}) shows allowed regions for various healing
lengths, all at a value of $n=10$. We see that as we move to smaller
values of $a_{0}$, the allowed bounds on $m$ and $V_{0}$ both move up,
as expected from equations (\ref{Vupper}) and (\ref{mupper}). More
physically, as the mass of the particle is increased, the repulsive
potential $V_{0}$ must increase to balance the stronger gravitational
force.

\section{Discussion}
\label{Discussion}

In this paper we have used techniques from condensed matter theory
in a cosmological setting to place bounds on parameters describing a
Dark Matter candidate, on the assumption that the Dark Matter halo
consists of a Bose-Einstein condensate, in which quantised vortices
reside. In the case of a laboratory BEC, self-gravitational forces
are not important but even in that case analytical progress is
limited. Using a simple physical argument however, we have shown
how rough limits on the consistency of such a model can be imposed.
Considering a Dark Matter particle of a particular mass, and a
vortex of a certain radius places constraints on the values that
the chemical potential and interaction potential can take. There
remain sizeable regions of parameter space in which the model
appears to be viable.

In future work, it would be interesting to investigate further
whether a Dark Matter candidate could reside in a coherent quantum
state, if the only interaction was gravitational. A less ambitious
undertaking would be to see if the Madelung transformation provides
a solution to the problem of defining all the relevant variables, as
suggested in Section \ref{Vortices in Gravitationally Coupled
BECs}. This would give a set of fluid equations that includes the
velocity giving rise to the stabilising centripetal force. One
problem to be anticipated in such a solution would be that the
velocity in the vortex core would still be ill-defined, as alluded
to in Section \ref{Bounds on Parameters}. The system would therefore
have to be solved by a more complete numerical method than we have
been able to implement so far.

We end with some comments on work that has been carried out since the original
version of this paper was written, although a complete review of the literature
is beyond the scope of this work.

First, vortices have been observed in turbulent interference patterns
generated in numerical simulations of the model we
discuss in this paper with no self-interaction; see, e.g. \cite{Schive:2014}.
In haloes described by this model, a random superposition of wave functions $\psi$ leads to regions of
phase-winding throughout the outer part of the halo; see, e.g. \cite{Lin:2018}. In condensed matter
this type of solution is sometimes called an incoherent soliton: the vortices are short-lived compared
to a timescale defined by the virial velocity, and surround volumes defined in scale by the de Broglie wavelength. Obviously these vortices do not form a regular distribution like a lattice, but are turbulent in nature.

Vortices in halo cores have been discussed by \cite{Rindler_Daller}; this work was expanded to complete studies of vortex lines in cores taking into account rotational motions, vertices in the outer halo and in cases with and without self-interaction, by \cite{aragon_mun} and \cite{Schobes}. Simulations in this field generally study the case with no self-interaction in the boson field. An exception is the work of \cite{Chen:2021} which considers a turbulent outer halo for both attractive and repulsive interactions, as long as these are not too strong.

\section*{Acknowledgements}

Mark Brook acknowledges support from the Science \& Technology
Facilities Council, and useful comments from Sean Carroll and David Tong. Most of
this work was completed when the authors were at the University of Nottingham, but the original
version of the paper, preprinted in 2009, was never published. This article is a revised
and updated version of that work. We thank the referees for very helpful and constructive comments.



\newpage

\bibliographystyle{mnras}
\bibliography{bibliography}

\begin{thebibliography}{}
\makeatletter
\relax
\def\mn@urlcharsother{\let\do\@makeother \do\$\do\&\do\#\do\^\do\_\do\%\do\~}
\def\mn@doi{\begingroup\mn@urlcharsother \@ifnextchar [ {\mn@doi@}
  {\mn@doi@[]}}
\def\mn@doi@[#1]#2{\def\@tempa{#1}\ifx\@tempa\@empty \href
  {http://dx.doi.org/#2} {doi:#2}\else \href {http://dx.doi.org/#2} {#1}\fi
  \endgroup}
\def\mn@eprint#1#2{\mn@eprint@#1:#2::\@nil}
\def\mn@eprint@arXiv#1{\href {http://arxiv.org/abs/#1} {{\tt arXiv:#1}}}
\def\mn@eprint@dblp#1{\href {http://dblp.uni-trier.de/rec/bibtex/#1.xml}
  {dblp:#1}}
\def\mn@eprint@#1:#2:#3:#4\@nil{\def\@tempa {#1}\def\@tempb {#2}\def\@tempc
  {#3}\ifx \@tempc \@empty \let \@tempc \@tempb \let \@tempb \@tempa \fi \ifx
  \@tempb \@empty \def\@tempb {arXiv}\fi \@ifundefined
  {mn@eprint@\@tempb}{\@tempb:\@tempc}{\expandafter \expandafter \csname
  mn@eprint@\@tempb\endcsname \expandafter{\@tempc}}}

\bibitem[\protect\citeauthoryear{{Arag{\'o}n-Mu{\~n}oz}, {Chac{\'o}n-Acosta}
  \& {Hernandez-Hernandez}}{{Arag{\'o}n-Mu{\~n}oz} et~al.}{2020}]{aragon_mun}
{Arag{\'o}n-Mu{\~n}oz} L.,  {Chac{\'o}n-Acosta} G.,   {Hernandez-Hernandez} H.,
   2020, \mn@doi [International Journal of Modern Physics B]
  {10.1142/S0217979220502719}, \href
  {https://ui.adsabs.harvard.edu/abs/2020IJMPB..3450271A} {34, 2050271}

\bibitem[\protect\citeauthoryear{{Arbey}, {Lesgourges}  \& {Salati}}{{Arbey}
  et~al.}{2003}]{Arbey:2003sj}
{Arbey} A.,  {Lesgourges} J.,   {Salati} P.,  2003, \prd, 68, 023511

\bibitem[\protect\citeauthoryear{{Berloff}}{{Berloff}}{2004}]{BerloffPade}
{Berloff} N.~G.,  2004, J. Phys. A, 37, 11729

\bibitem[\protect\citeauthoryear{{Boehmner} \& {Harko}}{{Boehmner} \&
  {Harko}}{2007}]{Boehmer:2007um}
{Boehmner} C.~G.,  {Harko} T.,  2007, \jcap, 0706

\bibitem[\protect\citeauthoryear{{Brook}}{{Brook}}{2010}]{Brook:2008PhD}
{Brook} M.,  2010, PhD thesis, University of Nottingham, UK

\bibitem[\protect\citeauthoryear{{Carroll}}{{Carroll}}{2007}]{Carroll:PC}
{Carroll} S.,  2007, Private Communication

\bibitem[\protect\citeauthoryear{{Carroll}}{{Carroll}}{2008}]{Carroll:CV}
{Carroll} S.,  2008, Gravity is an Important Force, \url
  {https://www.discovermagazine.com/the-sciences/gravity-is-an-important-force}

\bibitem[\protect\citeauthoryear{{Chen}, {Du}, {Lentz}, {Marsh}  \&
  {Niemeyer}}{{Chen} et~al.}{2021}]{Chen:2021}
{Chen} J.,  {Du} X.,  {Lentz} E.~W.,  {Marsh} D. J.~E.,   {Niemeyer} J.~C.,
  2021, \mn@doi [\prd] {10.1103/PhysRevD.104.083022}, \href
  {https://ui.adsabs.harvard.edu/abs/2021PhRvD.104h3022C} {104, 083022}

\bibitem[\protect\citeauthoryear{{Coddington}, {Engels}, {Schweikhard}  \&
  {Cornell}}{{Coddington} et~al.}{2003}]{Coddington:2003}
{Coddington} I.,  {Engels} P.,  {Schweikhard} V.,   {Cornell} E.~A.,  2003,
  \prl, 91, 100402

\bibitem[\protect\citeauthoryear{{Coles}}{{Coles}}{2002a}]{Coles:2002as}
{Coles} P.,  2002a, The Wave Mechanics of Large-scale Structure (\mn@eprint
  {arXiv} {astro-ph:0209576})

\bibitem[\protect\citeauthoryear{{Coles}}{{Coles}}{2002b}]{Coles:2001fw}
{Coles} P.,  2002b, \mnras, 330, 421

\bibitem[\protect\citeauthoryear{{Coles}}{{Coles}}{2005}]{Coles:2005yk}
{Coles} P.,  2005, \nat, 433, 248

\bibitem[\protect\citeauthoryear{{Coles} \& {Spencer}}{{Coles} \&
  {Spencer}}{2003}]{Coles:2002sj}
{Coles} P.,  {Spencer} K.,  2003, \mnras, 342, 176

\bibitem[\protect\citeauthoryear{{Feynman}}{{Feynman}}{1955}]{Feynman:1955}
{Feynman} R.~P.,  1955, Progress in Low Temperature Physics, 1, 17

\bibitem[\protect\citeauthoryear{{Goodman}}{{Goodman}}{2000}]{Goodman:2000tg}
{Goodman} J.,  2000, \mn@doi [New Astronomy] {10.1016/S1384-1076(00)00015-4},
  \href {https://ui.adsabs.harvard.edu/abs/2000NewA....5..103G} {5, 103}

\bibitem[\protect\citeauthoryear{{Guzman} \& {Urena-Lopez}}{{Guzman} \&
  {Urena-Lopez}}{2003}]{Guzman:2003kt}
{Guzman} F.~S.,  {Urena-Lopez} L.~A.,  2003, \prd, 68, 024023

\bibitem[\protect\citeauthoryear{{Hu}, {Barkana}  \& {Gruzinov}}{{Hu}
  et~al.}{2000}]{Hu:2000ke}
{Hu} W.,  {Barkana} R.,   {Gruzinov} A.,  2000, \prl, 85, 1158

\bibitem[\protect\citeauthoryear{{Jenkins} \& {the Virgo Consortium
  Collaboration}}{{Jenkins} \& {the Virgo Consortium
  Collaboration}}{1998}]{Jenkins:1997en}
{Jenkins} A.,  {the Virgo Consortium Collaboration} 1998, \apj, 499, 20

\bibitem[\protect\citeauthoryear{{Johnson} \& {Kamionkowski}}{{Johnson} \&
  {Kamionkowski}}{2008}]{JohnKam}
{Johnson} M.~C.,  {Kamionkowski} M.,  2008, \mn@doi [\prd]
  {10.1103/PhysRevD.78.063010}, \href
  {https://ui.adsabs.harvard.edu/abs/2008PhRvD..78f3010J} {78, 063010}

\bibitem[\protect\citeauthoryear{{Landau}}{{Landau}}{1941}]{Landau:1941}
{Landau} L.~D.,  1941, \prd, 60, 356

\bibitem[\protect\citeauthoryear{{Lesgourgues}, {Arbey}  \&
  {Salati}}{{Lesgourgues} et~al.}{2002}]{Lesgourgues:2002}
{Lesgourgues} J.,  {Arbey} A.,   {Salati} P.,  2002, New Astronomy Reviews, 46,
  761

\bibitem[\protect\citeauthoryear{{Lin}, {Schive}, {Wong}  \& {Chiueh}}{{Lin}
  et~al.}{2018}]{Lin:2018}
{Lin} S.-C.,  {Schive} H.-Y.,  {Wong} S.-K.,   {Chiueh} T.,  2018, \mn@doi
  [\prd] {10.1103/PhysRevD.97.103523}, \href
  {https://ui.adsabs.harvard.edu/abs/2018PhRvD..97j3523L} {97, 103523}

\bibitem[\protect\citeauthoryear{{Matos} \& {Guzman}}{{Matos} \&
  {Guzman}}{2000}]{Matos:1998vk}
{Matos} T.,  {Guzman} F.~S.,  2000, Class. Quantum Grav., 17, L9

\bibitem[\protect\citeauthoryear{{Mazur} \& {Mottola}}{{Mazur} \&
  {Mottola}}{2001}]{Mazur:2001fv}
{Mazur} P.~O.,  {Mottola} E.,  2001, arXiv e-prints, \href
  {https://ui.adsabs.harvard.edu/abs/2001gr.qc.....9035M} {pp gr--qc/0109035}

\bibitem[\protect\citeauthoryear{{Moore}, {Quinn}, {Governato}, {Stadel}  \&
  {Lake}}{{Moore} et~al.}{1999}]{Moore:1999gc}
{Moore} B.,  {Quinn} T.~R.,  {Governato} F.,  {Stadel} J.,   {Lake} G.,  1999,
  \mnras, 310, 1147

\bibitem[\protect\citeauthoryear{{Moroz}, {Penrose}  \& {Tod}}{{Moroz}
  et~al.}{1998}]{Moroz:1998dh}
{Moroz} I.~M.,  {Penrose} R.,   {Tod} P.,  1998, Class. Quantum Grav., 15, 2733

\bibitem[\protect\citeauthoryear{{Navarro}, {Frenk}  \& {White}}{{Navarro}
  et~al.}{1997}]{Navarro:1996gj}
{Navarro} J.,  {Frenk} C.~S.,   {White} S. D.~M.,  1997, \apj, 490, 493

\bibitem[\protect\citeauthoryear{{Onsager}}{{Onsager}}{1949}]{Onsager:1949}
{Onsager} L.,  1949, Nuovo Cimento, 6, 249

\bibitem[\protect\citeauthoryear{{Osbourne}}{{Osbourne}}{1950}]{Osbourne:1950}
{Osbourne} D.~V.,  1950, Proc. Phys. Soc. A, 63, 909

\bibitem[\protect\citeauthoryear{{Packard} \& {Sanders Jr}}{{Packard} \&
  {Sanders Jr}}{1972}]{Packard:1972}
{Packard} R.~E.,  {Sanders Jr} T.~M.,  1972, Phys. Rev. A., 6, 799

\bibitem[\protect\citeauthoryear{{Peebles}}{{Peebles}}{2000}]{Peebles:2000yy}
{Peebles} P. J.~E.,  2000, \apj Lett., 534, L127

\bibitem[\protect\citeauthoryear{{Pethick} \& {Smith}}{{Pethick} \&
  {Smith}}{2008}]{Pethick:2008}
{Pethick} C.~J.,  {Smith} H.,  2008, Bose-Einstein Condensation in Dilute
  Gases, 2nd edn.
Cambridge University Press

\bibitem[\protect\citeauthoryear{{Rindler-Daller} \&
  {Shapiro}}{{Rindler-Daller} \& {Shapiro}}{2012}]{Rindler_Daller}
{Rindler-Daller} T.,  {Shapiro} P.~R.,  2012, \mn@doi [\mnras]
  {10.1111/j.1365-2966.2012.20588.x}, \href
  {https://ui.adsabs.harvard.edu/abs/2012MNRAS.422..135R} {422, 135}

\bibitem[\protect\citeauthoryear{{Roberts} \& {Berloff}}{{Roberts} \&
  {Berloff}}{2001}]{RobertsBerloff}
{Roberts} P.~H.,  {Berloff} N.~G.,  2001, in {Barenghi} C.,  {Donnelly} R.~J.,
   {Vinen} R.~F.,  eds,  Lecture Notes in Physics Vol. 571, Quantized Vortex
  Dynamics and Superfluid Turbulence. Springer-Verlag, pp 235--257

\bibitem[\protect\citeauthoryear{{Romanowsky}, {Douglas}, {Arnaboldi},
  {Kuijken}, {Merrifield}, {Napolitano}, {Capaccioli}  \&
  {Freeman}}{{Romanowsky} et~al.}{2003}]{Romanowsky:2003qv}
{Romanowsky} A.~J.,  {Douglas} N.~G.,  {Arnaboldi} M.,  {Kuijken} K.,
  {Merrifield} M.~R.,  {Napolitano} N.~R.,  {Capaccioli} M.,   {Freeman} K.~C.,
   2003, \mn@doi [Science] {10.1126/science.1087441}, \href
  {https://ui.adsabs.harvard.edu/abs/2003Sci...301.1696R} {301, 1696}

\bibitem[\protect\citeauthoryear{{Sadd}, {Chester}  \& {Reatto}}{{Sadd}
  et~al.}{1997}]{Sadd:1997sv}
{Sadd} M.,  {Chester} G.~V.,   {Reatto} L.,  1997, \prl, 79, 2490

\bibitem[\protect\citeauthoryear{{Schive}, {Chiueh}  \& {Broadhurst}}{{Schive}
  et~al.}{2014}]{Schive:2014}
{Schive} H.-Y.,  {Chiueh} T.,   {Broadhurst} T.,  2014, \mn@doi [Nature
  Physics] {10.1038/nphys2996}, \href
  {https://ui.adsabs.harvard.edu/abs/2014NatPh..10..496S} {10, 496}

\bibitem[\protect\citeauthoryear{{Schobesberger}, {Rindler-Daller}  \&
  {Shapiro}}{{Schobesberger} et~al.}{2021}]{Schobes}
{Schobesberger} S.~O.,  {Rindler-Daller} T.,   {Shapiro} P.~R.,  2021, \mn@doi
  [\mnras] {10.1093/mnras/stab1153}, \href
  {https://ui.adsabs.harvard.edu/abs/2021MNRAS.505..802S} {505, 802}

\bibitem[\protect\citeauthoryear{{Seidel} \& {Suen}}{{Seidel} \&
  {Suen}}{1990}]{Seidel:1990jh}
{Seidel} E.,  {Suen} W.~M.,  1990, \prd, 42, 384

\bibitem[\protect\citeauthoryear{{Seidel} \& {Suen}}{{Seidel} \&
  {Suen}}{1991}]{Seidel:1991zh}
{Seidel} E.,  {Suen} W.~M.,  1991, \prl, 66, 1659

\bibitem[\protect\citeauthoryear{{Short}}{{Short}}{2007}]{Short:2007PhD}
{Short} C.~J.,  2007, PhD thesis, University of Nottingham, UK

\bibitem[\protect\citeauthoryear{{Short} \& {Coles}}{{Short} \&
  {Coles}}{2006}]{Short:2006md}
{Short} C.~J.,  {Coles} P.,  2006, \jcap, 0612, 012

\bibitem[\protect\citeauthoryear{{Silverman} \& {Mallett}}{{Silverman} \&
  {Mallett}}{2002}]{Silverman:2002qx}
{Silverman} M.~P.,  {Mallett} R.~L.,  2002, Gen. Rel. Grav., 34, 633

\bibitem[\protect\citeauthoryear{{Spiegel}}{{Spiegel}}{1980}]{Spiegel:1980fd}
{Spiegel} E.~A.,  1980, Physica D, 1, 236

\bibitem[\protect\citeauthoryear{{Uson}, {Boughn}  \& {Kuhn}}{{Uson}
  et~al.}{1990}]{Uson:1990}
{Uson} J.~M.,  {Boughn} S.~P.,   {Kuhn} J.~R.,  1990, Science, 250, 539

\bibitem[\protect\citeauthoryear{{Vilenkin} \& {Shellard}}{{Vilenkin} \&
  {Shellard}}{1986}]{Vilenkin:1986hg}
{Vilenkin} A.,  {Shellard} E. P.~S.,  1986, Cosmic Strings and Other
  Topological Defects.
Cambridge University Press

\bibitem[\protect\citeauthoryear{{Vinen}}{{Vinen}}{2000}]{Vinen:2000ts}
{Vinen} W.~F.,  2000, J. Low. Temp. Phys., 121

\bibitem[\protect\citeauthoryear{{Widrow} \& {Kaiser}}{{Widrow} \&
  {Kaiser}}{1993}]{Widrow:1993qq}
{Widrow} L.~M.,  {Kaiser} N.,  1993, \apj Lett., 416, L71

\bibitem[\protect\citeauthoryear{{Yu} \& {Morgan}}{{Yu} \&
  {Morgan}}{2002}]{Yu:2002sz}
{Yu} R.~P.,  {Morgan} M.~J.,  2002, Class. Quantum Grav., 19, L157

\makeatother
\end{thebibliography}

\end{document}